\documentclass[aps,prc,twocolumn,superscriptaddress,nofootinbib]{revtex4}
\usepackage{graphicx,amsfonts,color,epsfig,epstopdf}
\usepackage{footnote,multirow}
\usepackage{amsmath}
\usepackage{amssymb}
\usepackage{physics}
\usepackage[utf8]{inputenc}

\newcommand{\Nmax}{$N_\mathrm{max}$ }
\newcommand{\gs}{ground state }

\newcommand{\SU}[1]{\ensuremath{\mathrm{SU}( #1 )}}

\newcommand{\SO}[1]{\ensuremath{\mathrm{SO}( #1 )}}

\newcommand{\SpR}[1]{\ensuremath{\mathrm{Sp}( #1,\mathbb{R} )}}
\newcommand{\CG}[3]{\ensuremath{\langle#1;#2|\,#3\rangle}}

\newcommand{\RedCG}[3]{\ensuremath{\langle#1;#2\|#3\rangle}}
\newcommand{\Wigsixj}[6]{\ensuremath{ \left\{ \begin{matrix}#1 & #2 & #3\cr
#4 & #5 & #6 \end{matrix} \right\} }}
\newcommand{\Wigninej}[9]{\ensuremath{\left\{\begin{matrix}#1 & #2 & #3\cr
#4 & #5 & #6 \cr #7 & #8 & #9 \end{matrix}\right\}}}

\newcommand{\braketop}[3]{\ensuremath{\left\langle #1 | #2 | #3 \right\rangle}}

\newcommand{\expV}[1]{\ensuremath{\left\langle #1 \right\rangle}}
\newcommand{\RedCGw}[4]{\ensuremath{\langle\omega_{#1} \kappa_{#1} L_{#1};\omega_{#2} \kappa_{#2} L_{#2}\|\omega_{#3} \kappa_{#3} L_{#3}\rangle}_{\rho_{#4}} }

\newcommand{\half}{\ensuremath{\textstyle{\frac{1}{2}}}}

\begin{document}
\title{SU(3)-guided Realistic Nucleon-nucleon Interactions for Large-scale Calculations}
\author{G. H. Sargsyan}
\affiliation{Department of Physics and Astronomy, Louisiana State University, Baton Rouge, LA 70803, USA}
\author{K. D. Launey}
\affiliation{Department of Physics and Astronomy, Louisiana State University, Baton Rouge, LA 70803, USA}
\author{R. B. Baker}
\affiliation{Department of Physics and Astronomy, Louisiana State University, Baton Rouge, LA 70803, USA}
\author{T. Dytrych}
\affiliation{Department of Physics and Astronomy, Louisiana State University, Baton Rouge, LA 70803, USA}
\affiliation{Nuclear Physics Institute, 250 68 Rez, Czech Republic}
\author{J. P. Draayer}
\affiliation{Department of Physics and Astronomy, Louisiana State University, Baton Rouge, LA 70803, USA}

\begin{abstract}

We examine nucleon-nucleon realistic interactions, based on their SU(3) decomposition to SU(3)-symmetric components. We  find that many of these interaction components are negligible, which, in turn,  allows us to identify a subset of physically relevant  components that are sufficient to describe the structure of low-lying states in $^{12}$C and related observables, such as excitation energies, electric quadrupole transitions and rms radii. We find that paring the interaction down to half of the  SU(3)-symmetric components or more  yields results that practically  coincide with the  corresponding \textit{ab initio} calculations with the full interaction. In addition, we show that while various realistic interactions differ in their SU(3) decomposition, their renormalized effective counterparts exhibit a striking similarity and composition that can be linked to dominant nuclear features such as deformation, pairing, clustering, and spin-orbit effect.

\end{abstract}

\maketitle
\section{Introduction}
{\it Ab initio} calculations aim to describe nuclear features while employing high-precision interactions that describe two- and three-nucleon systems (often referred to as ``realistic interactions"),  such as those derived from meson
exchange theory \cite{machleidt1987bonn, machleidt1989meson} (e.g. CD-Bonn \cite{Machleidt01}), chiral effective field theory \cite{van1994few, epelbaum2009modern, machleidt2011chiral} (e.g. NNLO$_\mathrm{opt}$ \cite{Ekstrom13} and N3LO \cite{EntemM03}), or $J$-matrix inverse scattering    (JISP16 \cite{ShirokovMZVW07, Shirokov2010nn}). As such calculations do not depend on any information about the nucleus in consideration,  these methods can be used in nuclear regions where experimental data is currently sparse or not available, e.g., along the pathways of nucleosynthesis and toward a further exploration of exotic physics of rare isotopes.

While realistic interactions build upon rich physics at the nucleon-nucleon (NN) level, it is impossible to identify terms in the interaction that are responsible for emergent dominant features in nuclei, such as deformation, pairing, and clustering. These features, which are revealed in even the earliest of data on nuclear structure, have informed many successful nuclear models such as Elliott's \SU{3} model \cite{Elliott58, Elliott58b,ElliottH62} and Bohr collective model \cite{BohrMottelson69}  with a focus on deformation, as well as algebraic \cite{Racah42,Belyaev58} and exact \cite{Richardson1964} pairing models. 
Recently, we have shown that calculations that consider Hamiltonians that build upon the ones used in these earlier studies and, in addition, allow for configuration mixing \cite{DreyfussLTDB13,TobinFLDDB14,MioraLKPD2019}, yield results that are consistent with the ones in the {\it ab initio} symmetry-adapted no-core shell model (SA-NCSM) \cite{LauneyDD16, DytrychLDRWRBB20}. In particular, the no-core symplectic model (NCSpM) has offered  successful descriptions for excitation energies,  monopole and quadrupole transitions, quadrupole moments, and rms radii for  a range of nuclei (from $A$=8 to $A$=24 systems, including cluster effects in the $^{12}$C Hoyle state) \cite{DreyfussLTDB13,TobinFLDDB14,DreyfussLTDBDB17}, by employing quadrupole-quadrupole ($Q \cdot Q$) and spin-orbit interaction terms. In Ref. \cite{MioraLKPD2019}, exact solutions to the  shell model plus isoscalar and isovector pairing have been provided for low-lying $0^+$ states and, e.g.,  the energy of the lowest isobaric analog state in  $^{12}$C  has been shown to agree with the corresponding \textit{ab initio} findings. Therefore, it is interesting to trace this similarity in outcomes down to specific features of the realistic interactions.

In this paper, we provide new insight into correlations within realistic interactions through the use of the deformation-related \SU{3} symmetry. Specifically,
we show  that only a part of the nucleon-nucleon interaction appears
to be essential for the description of nuclear dynamics, especially at low energies. When expressed in the \SU{3} symmetry-adapted basis, the interaction -- given as \SU{3} tensors -- shows a clear preference toward a specific subset of tensors, allowing us to determine its dominant components. 
Most importantly, these features appear regardless of the underlying theory used to construct the interaction. Furthermore, an almost universal behavior is revealed by  ``soft-core" potentials such as JISP16, or by the renormalized (``softened'') counterparts of ``harder" interactions that use, e.g., Okubo-Lee-Suzuki (OLS) \cite{Okubo1954diagonalization, LeeSuzuki80} and  Similarity Renormalization Group (SRG) \cite{BognerFP07} renormalization techniques. And further, to complete the picture, we show that these features are directly linked to the important physics, i.e., deformation, clustering, pairing, and spin-orbit effects, that drove the development of earlier, and considerably simpler, schematic models.

The importance of various 
 interaction components is studied in
 SA-NCSM calculations. In particular, we study nuclear structure observables of $^{12}$C, such as the low-lying excitation spectrum, B(E2) reduced transition probabilities  and root mean square (rms) radii. We compare the results that use the entire interaction with those that use interactions that have been selected down to their dominant components. The agreement observed for all these observables is remarkable, even when a small fraction of the interaction is used.

\section{Theoretical method}

\begin{figure*} [t]
\centering
\includegraphics[width=0.99\textwidth]{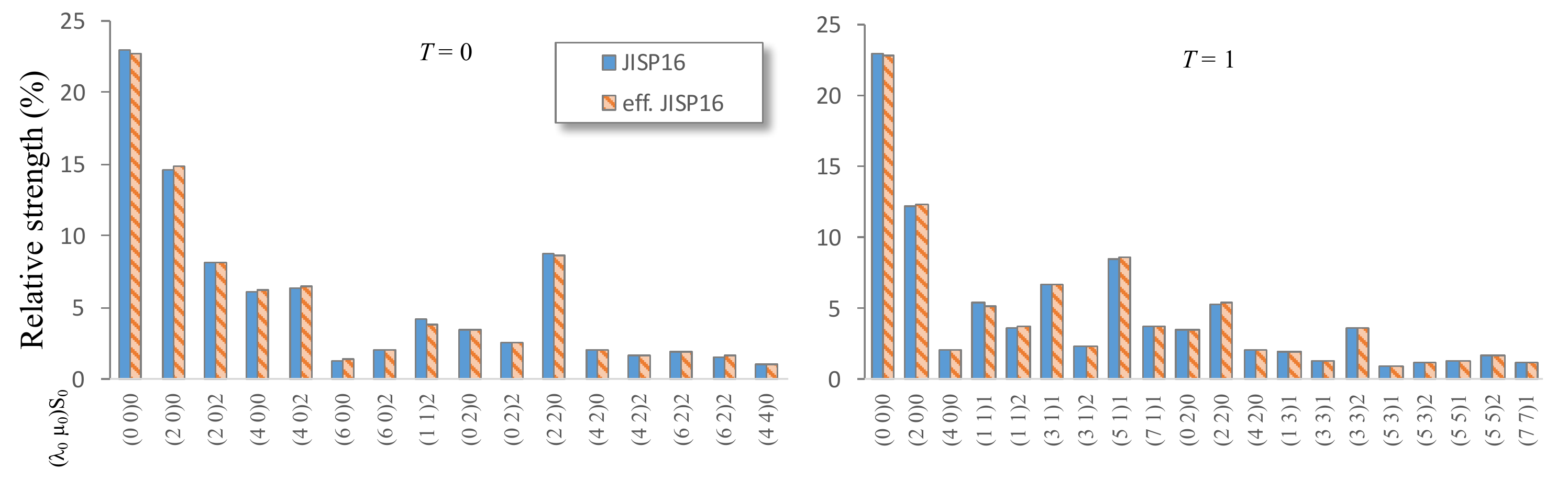}
\includegraphics[width=0.99\textwidth]{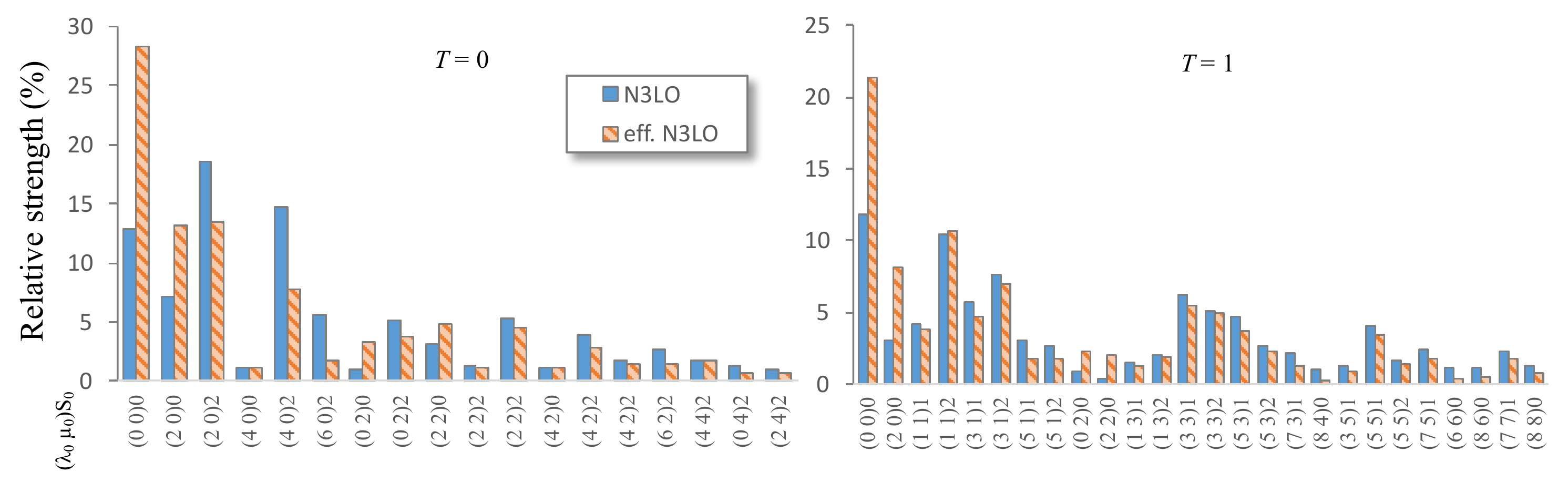}
\caption{\label{fig:decomp} Relative strengths $\mathfrak s$ (in \%) for the \SU{3}-coupled JISP16 (top) and N3LO (bottom) NN interactions and their effective counterparts with $\hbar\Omega=15$ MeV and 20 MeV, respectively, in the \Nmax = 6 model space. The “eff. JISP16” is obtained by the OLS technique for $A$=12, while “eff. N3LO” is by SRG with $\lambda_\mathrm{SRG}$ = 2.0 fm$^{-1}$. $T$ is the isospin of the two nucleon system.
A set of $(\lambda_0 \mu_0)S_0$ quantum numbers and its conjugate correspond to each of the interaction terms.  Only terms with $>$1\% relative strength for each $T$ are shown; there are more than 120 terms with less than 1\% strength for this model space. 
} 
\end{figure*}

\subsection{SA-NCSM framework}
The SA-NCSM is a no-core shell model with an \SU{3}-coupled or \SpR{3}-coupled symmetry-adapted basis \cite{LauneyDD16,DytrychLDRWRBB20}. Similar to NCSM \cite{NavratilVB00,NavratilVB00b}, it uses a harmonic oscillator (HO) basis,
where the HO major shells are separated by a parameter
$\hbar \Omega$. 
The model space is capped by an \Nmax cutoff which is the maximum total number of oscillator quanta above the lowest HO configuration for a given nucleus. The SA-NCSM calculates eigenvalues and eigenvectors of the nuclear interaction Hamiltonian and subsequently uses the eigenvectors for calculations of the nuclear observables. The 
results approach the exact value as the \Nmax increases, and at  the \Nmax $\rightarrow \infty$ limit they become independent of the HO parameter $\hbar \Omega$.
Within a given complete \Nmax model space, the SA-NCSM results exactly match those of the NCSM for the same interaction. The use of symmetries in SA-NCSM allows one to select the model space by considering only the physically relevant subspace, which is only a fraction of the corresponding  complete \Nmax space.

In the SA-NCSM, the SA basis is constructed using an efficient group-theoretical algorithm for each HO major shell \cite{DraayerLPL89}.
While we do not use explicit construction of conventional NCSM bases, for completeness, we show the unitary transformation from a two-particle $JT$-coupled basis state to an \SU{3}-coupled state:
\begin{eqnarray}
&&\ket{ \eta_r \eta_s \omega \kappa (LS) \Gamma M_\Gamma } \nonumber \\
&=&\frac{1}{ \sqrt{1+\delta_{\eta_r\eta_s}} }
\{a^\dagger_{(\eta_r\, 0) \half  } \times a^\dagger_{(\eta_s\,0) \half } \}^{ \omega \kappa (LS) \Gamma M_\Gamma}\ket{0} \nonumber \\
&=&\frac{1}{ \sqrt{1+\delta_{\eta_r\eta_s}} }
\sum_{\substack{ l_r l_s\\ j_r j_s}}\Pi_{j_r j_s L S} \RedCG{(\eta_r\ 0)l_r}{(\eta_s\ 0) l_s}{\omega \kappa L} \nonumber \\
&\times&\Wigninej{l_r}{l_s}{L}{1/2}{1/2}{S}{j_r}{j_s}{J} \{a^\dagger_{r} \times a^\dagger_{s} \}^{  \Gamma M_\Gamma}\ket{0},
\label{su3basis}
\end{eqnarray}
where we use conventional labels $\textstyle { r(s)=\{\eta (l\,\frac{1}{2})j \textstyle{ t=\frac{1}{2}}\} }$ and $\Gamma=JT$, with $\eta=0,1,2,\dots$ is the oscillator shell number and $\Pi_j=\sqrt{2j+1}$, and with $a^\dagger_{(\eta \,0)\half}$ being the creation operator that creates a particle of spin $\half$ and in a HO major shell $\eta$.
We use \SU{3} quantum numbers, $\textstyle { \omega \equiv (\lambda\, \mu) = (\eta_r\, 0) \times (\eta_s\, 0) }, \, \tilde{\omega} \equiv (\mu\, \lambda) $, and $\kappa$ the multiplicity of total orbital momentum $L$ for a given $\omega$; $S$ is the total intrinsic spin, and $\RedCG{}{}{}$ are reduced SU(3) Clebsch-Gordan coefficients.

\subsection{\SU{3} interaction tensors}

Two-body isoscalar (charge-independent) interactions are typically given in a representation of a $JT$-coupled HO basis, $\ket{rs\Gamma M_\Gamma}$,
that is, $V^{\Gamma}_{rstu} = \braketop{rs \Gamma M_\Gamma=0}{V}{tu \Gamma M_\Gamma=0}$. This takes advantage of the fact that this interaction transforms as a scalar under rotations in coordinate and isospin space, that is, it  is an \SO{3}$\times$ \SU{2}$_T$ tensor of rank zero. Analogously, the interaction can be represented in an $\SU{3}\times$\SU{2}$_S\times$\SU{2}$_T$-coupled HO basis 
$\ket{ \eta_r \eta_s \omega \kappa (LS) \Gamma M_\Gamma }$  (\ref{su3basis}).
The corresponding interaction matrix elements 
are similarly given as
$V^{\Gamma}_{(\chi \omega \kappa L S)_{fi}}\equiv  \braketop{(\chi \omega \kappa (LS) \Gamma M)_f}{V}{(\chi \omega \kappa (LS) \Gamma M)_i}$,
 with $\chi \equiv \{\eta_r \eta_s\}$ and with symmetry properties $V^{\Gamma}_{(\chi \omega \kappa L S)_{if}}=V^{\Gamma}_{(\chi \omega \kappa L S)_{fi}}$. Using that the interaction can be represented as a sum of \SU{3}$\times $\SU{2}$_S$ tensors, $V=\sum_{\rho_0 \omega_0 \kappa_0 S_0}V^{\rho_0 \omega_0 \kappa_0 S_0}$, the matrix elements can be further reduced with respect to \SU{3} and the spin-isospin space (for $T_0=0$),
 $V_{(\chi \omega S)_{if};T}^{\rho_0 \omega_0 \kappa_0 S_0}\equiv \braketop{(\chi \omega S)_f;T}{|V^{\omega_0 \kappa_0 S_0}|}{(\chi \omega S)_i;T}_{\rho_0}$
 (see Appendix).

The following conjugation relations hold for the $\SU{3}\times\SU{2}_S$ tensors,
\begin{eqnarray}
V_{(\chi \omega S)_{if};T}^{\rho_0 \omega_0 \kappa_0 S_0}&=&(-)^{S_i-S_f+S_0}(-)^{\omega_f-\omega_i}\sqrt{\frac{\dim \omega_f}{\dim \omega_i}}
V_{(\chi \omega S)_{fi};T}^{\rho_0 \tilde \omega_0 \kappa_0 S_0} \nonumber \\
V_{(\chi \omega S)_{ii};T}^{\rho_0 \omega_0 \kappa_0 S_0}&=&(-)^{S_0}V_{(\chi \omega S)_{ii};T}^{\rho_0  \tilde \omega_0 \kappa_0 S_0}, 
\end{eqnarray}
 where 
 \begin{equation}
 \dim \omega = \frac{1}{2}(\lambda +1)( \mu+1)(\lambda+\mu + 2)
 \label{dim}.
 \end{equation}

 To simplify the equations in the paper, we introduce a symmetrized tensor,
\begin{equation}
v_{(\chi \omega S)_{if};T}^{\rho_0 \omega_0 \kappa_0 S_0}=(-)^{\omega_i-S_i-T }\sqrt{\dim \omega_i}V_{(\chi \omega S)_{if};T}^{\rho_0 \omega_0 \kappa_0 S_0},
\end{equation}
with a conjugation relation,
\begin{equation}
v_{(\chi \omega S)_{if};T}^{\rho_0 \omega_0 \kappa_0 S_0}=(-)^{S_0}v_{(\chi \omega S)_{fi};T}^{\rho_0 \tilde \omega_0 \kappa_0 S_0}.
\end{equation}
We note that, in the case when $\chi_i=\chi_f$, $\omega _i =\omega _f$, and $S_i = S_f$, we will use the notation $v_{(\chi \omega S);T}^{\rho_0 \omega_0 \kappa_0 S_0}$.

\subsection{Strength of \SU{3}  interaction tensors}

\begin{figure}[t]
\includegraphics[width=0.48\textwidth]{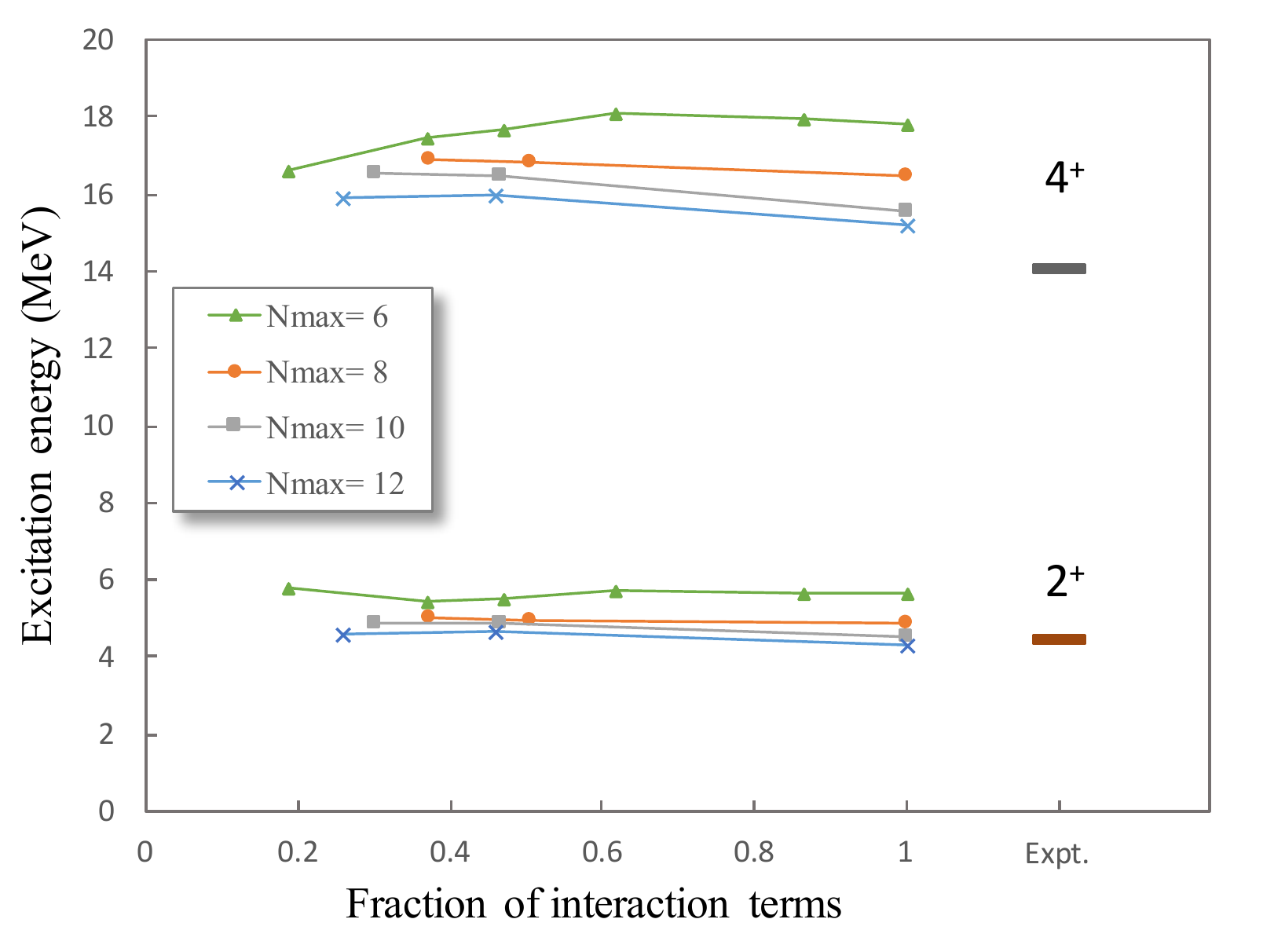}
\caption{\label{fig:12Cen} Excitation energy of the first $2^+$ and $4^+$ states in $^{12}$C from SA-NCSM calculations (connected lines) as a function of the fraction of the terms kept in the interaction, and compared to experiment \cite{ASelove90} (labeled as ``Expt.").  Results for \Nmax = 6, 8, 10, and 12 are shown for various selections of the JISP16 interaction with $\hbar\Omega=15$ MeV.  Specifically, the value 1 on the abscissa indicates the full interaction (100\%) was used, while an abscissa value of 0.4 implies that only the most significant 40\% of the tensors were retained, etc. }
\end{figure}


The significance of the various \SU{3} tensors can be estimated by their Hilbert-Schmidt norm, which is analogous to the norm of a matrix $A$ defined as $||A||=\sqrt[]{\sum_{ij}A_{ij}A_{ji}}$. In particular, the strength of a Hamiltonian $H$ can be estimated by  the norm $\sigma_H$ constructed as \cite{HechtD74,French66, FrenchR71, ChangFT71, KotaH10, LauneyCPC2014}
\begin{equation}
\sigma^2_H=
\expV{(H- \expV{H})^\dagger (H- \expV{H}) }=\expV{H^2 }-\expV{H}^2,
\label{sigma}
\end{equation}
 where $\expV{\dots}\equiv \frac{1}{  \mathcal N } \text{Tr}(\dots)$ specifies the trace of the Hamiltonian matrix divided by the  ${\mathcal N}$ number of diagonal matrix elements. In the present study, $H$ 
 is a two-body Hamiltonian, and  ${\mathcal N}$ enumerates all possible two-particle configurations.

\begin{figure*}
\includegraphics[width=1 \textwidth]{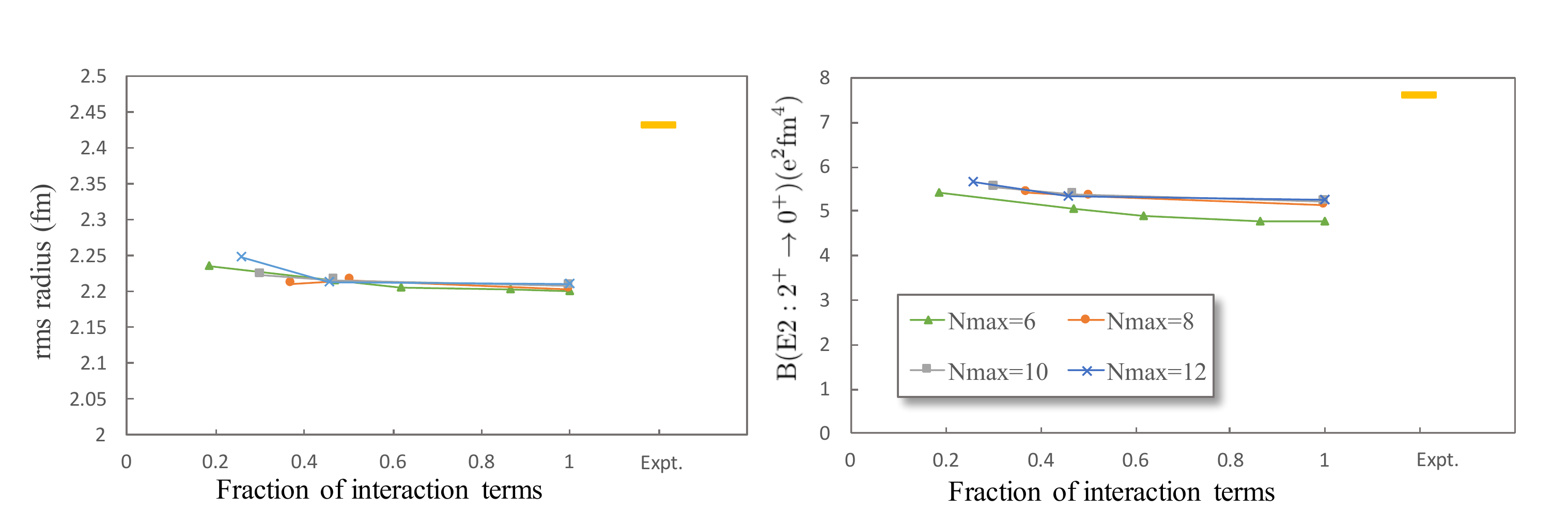}
\caption{\label{fig:obs} Same as Fig. \ref{fig:12Cen}, but for the rms radius (in fm) of the $^{12}$C \gs (experimental value from Ref. \cite{Tanihata85}) and the B(E2: $2^+_1 \rightarrow 0^+_1$) value  (in e$^2$fm$^4$) (experimental value from \cite{ASelove90}) as a function of the fraction of the terms kept in the interaction. SA-NCSM calculations use various selections for the JISP16 interaction  for $\hbar\Omega=15$ MeV and different \Nmax model spaces. 
}
\end{figure*}

For given $T_f=T_i=T$ and a $\ket{ \chi^* \omega \kappa (LS) \Gamma M_\Gamma }$ basis with $\chi ^* \equiv \{ \eta_r \eta_s \}, \eta_r \leq \eta_s $, the norm $\sigma_{\omega_0 \kappa_0 S_0;T}$ of each SU(3)-symmetric tensor is determined using Eq. (\ref{sigma}):
\begin{eqnarray}
\sigma^2_{\omega_0 \kappa_0 S_0;T}&=&\frac{1}{  \mathcal N  }\sum_{(\chi^*\omega S)_{f,i}\rho_0} \frac{1}{\Pi^2_{T_f S_0 T_0}\dim \omega_0}
| v_{(\chi \omega S)_{if};T}^{\rho_0 \omega_0 \kappa_0 S_0} |^2 \nonumber \\
&&-(V_c^{\omega_0 \kappa_0 S_0;T})^2,
\end{eqnarray}
where the number of two-particle basis states ${\mathcal N}$ and the  average monopole part $V_c^{\omega_0 \kappa_0 S_0}=\expV{V^{\omega_0 \kappa_0 (L_0=S_0 S_0)\Gamma_0=0 M_{\Gamma_0} =0}}$ are given, respectively, as
\begin{equation}
\mathcal N = \sum_{\chi^*\omega \kappa  LSJ M_J}  1 = \sum_{\chi^*\omega S} \Pi^2_{S} \dim \omega,
\end{equation}

\begin{eqnarray}
V_c^{\omega_0 \kappa_0 S_0} 
=\frac{1}{ \mathcal N} \sum_{\substack{\chi^*\omega \kappa \\ LSJ \rho_0}} \frac{\Pi^2_J\Pi_L}{\Pi_{S_0T}\sqrt{{\rm dim}\omega}} (-1)^{S_0+L+J-T-\omega}  \nonumber \\
\times \Wigsixj{L}{S}{J}{S}{L}{S_0}
\langle\omega \kappa L;\omega_{0} \kappa_{0} L_{0}\|\omega \kappa L \rangle_{\rho_0} 
v_{(\chi \omega S);T}^{\rho_0 \omega_0 \kappa_0 S_0}. 
\end{eqnarray}

For a given isospin $T$, the strength of the entire Hamiltonian $H_T$ is determined by the strengths of its components, $\sigma^2_{H_T}=\sum_{\omega_0 \kappa_0 S_0} \sigma^2_{\omega_0 \kappa_0 S_0;T}$. We can then define a relative strength for each SU(3)-symmetric component ($\omega_0 \kappa_0 S_0$)  as 
\begin{equation}
{\mathfrak s}^2_{\omega_0 \kappa_0 S_0;T}=\frac{\sigma^2_{\omega_0 \kappa_0 S_0;T} }{\sigma^2_{H_T}}=\frac{\sigma^2_{\omega_0 \kappa_0 S_0;T} }{\sum_{\omega_0 \kappa_0 S_0}{\sigma^2_{\omega_0 \kappa_0 S_0;T}}}.
\label{relsigma}
\end{equation}

Using Eq. (\ref{JTtoSU3}),
we can decompose any two-body interaction into \SU{3}-symmetric components.  The contribution of each of the components within the interaction is given by its relative strength (\ref{relsigma}) (see Fig. \ref{fig:decomp} for the realistic JISP16 and N3LO interactions). As can be seen from these results, only a small number of \SU{3} tensors dominate the interaction, with the vast majority of the components having less than 1\% of the total strength. Similar behavior is observed  for other interactions. It should be noted that in the $JT$-coupled basis, no such dominance of interaction matrix elements is apparent. This exercise demonstrates a long-standing principle that holds across all of physics; namely, one should work within a framework that is as closely aligned with the dynamics as possible.

\section{Results and Discussions}

\subsection{Observables in $^{12}$C}

The decomposition of the interaction in the \SU{3} basis allows us to choose sets of major components to construct new selected interactions. These interactions can be used for calculations of various nuclear properties that can then be compared to the results from the initial interaction. 
In this way, we can examine how sensitive specific nuclear properties are to the interaction components. 

Several selected interactions were constructed for this study. The selection is done by ordering the interaction tensors from the highest relative strength to the lowest and then including the largest ones to add up to 60 - 90\% of the initial total strength. Depending on the \Nmax of the interaction the number of selected \SU{3} tensors differs. For example, JISP16 interaction in \Nmax = 10, $\hbar\Omega$=15 MeV has overall 169 unique $(\lambda_0 \mu_0)S_0$ tensors, out of which 51 largest ones account for about 80\% of the total strength. After selection the total strengths are not rescaled to the initial interaction. Throughout this work we will refer to selected interactions  in terms of the fraction of interaction tensors kept, that is the number of SU(3)-symmetric components in the selected interaction relative to the number of all such components in  the initial interaction for a given \Nmax and $\hbar\Omega$.  

Analysis of the results shows that low-lying excitation energies of $^{12}$C are not sensitive to the number of selected \SU{3} tensors, given that the most dominant ones are included in the interaction (Fig. \ref{fig:12Cen}). With only half of the interaction tensors the excitation energies essentially do not differ from the corresponding  results that use the full interaction, and even with less than 30\% of the interaction components the deviation for most of the values is  insignificant. The comparatively large deviation in 4$^+$ energy for \Nmax = 6 that happens when about 20\% of the \SU{3} components are used is likely due to the small model space. This issue disappears in higher \Nmax values, and even \Nmax = 6 results for the 2$^+$ state compare remarkably well to the initial interaction for all selections.

The selected interactions yield very close results to the initial one for other observables as well. For example, the $^{12}$C rms radius of the ground state and the B(E2: $2^+ \rightarrow 0^+$) have very low dependence on the selection (Fig. \ref{fig:obs}), with variations nearly inconsequential compared to the deviations from the experiment (the underprediction of these observables for the
JISP16 interaction has been addressed, e.g., in Ref. \cite{DytrychMLDVCLCS11}). Specifically, the values are essentially the same when half of the interaction components are used. With less than 30\% of interaction components, the difference from the initial interaction results is less than 2\% for rms radius and less than 7\% for B(E2). Thus, small deviations start to appear only at significantly trimmed  interactions, indicating that the long-range physics is mostly preserved when only the dominant interaction terms are used. 

\begin{figure} [b]
\includegraphics[width=0.5 \textwidth]{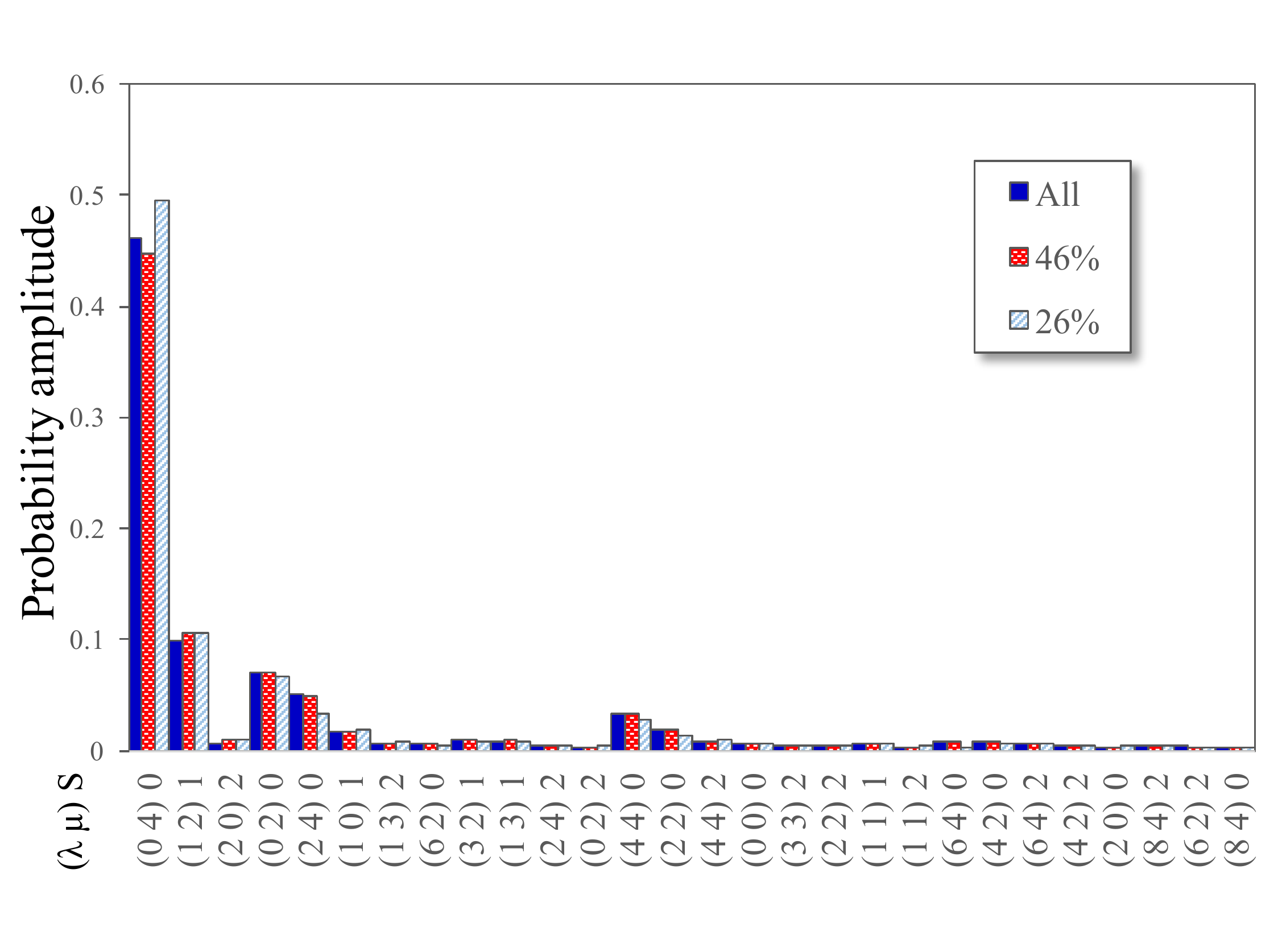}
\caption{\label{fig:wfns} Probability amplitudes for the $(\lambda\,\mu)S$  configurations  that make up the $^{12}$C  \gs ($0^+_1$), calculated in \Nmax= 12 model space using JISP16 interaction for $\hbar\Omega=15$ MeV (labeled by ``All") and two selected interactions (labeled by the fraction of the interaction components kept, 46\% and 26\%).
Only states with probability amplitudes  $>0.003$ are shown.
}
\end{figure}

In addition, vital information about the nuclear structure can be found through analysis of the  ($\lambda \mu$)$S$ configurations that comprise the SA-NCSM wavefunction. This
uncovers the physically relevant features that arise from the complex nuclear dynamics as shown in Ref. \cite{LauneyDD16}. In other words, the wavefunctions 
contain a manageable number of major \SU{3} components that account for most of the underlying physics. Indeed, we find that calculations with various selected interactions largely preserve the major components of the wavefunction (Fig. \ref{fig:wfns}). For the \gs of $^{12}$C calculated in the \Nmax= 12 model space the probability amplitude for each set of the quantum numbers ($\lambda \mu$)$S$ almost does not change when a little less then half (46\%) of the JISP16 interaction tensors are used for the calculations. Even with about quarter (26\%) of the tensors, the \SU{3} structure remains the same with only a slight difference in the amplitudes.

\begin{figure} 
\includegraphics[width=0.45 \textwidth]{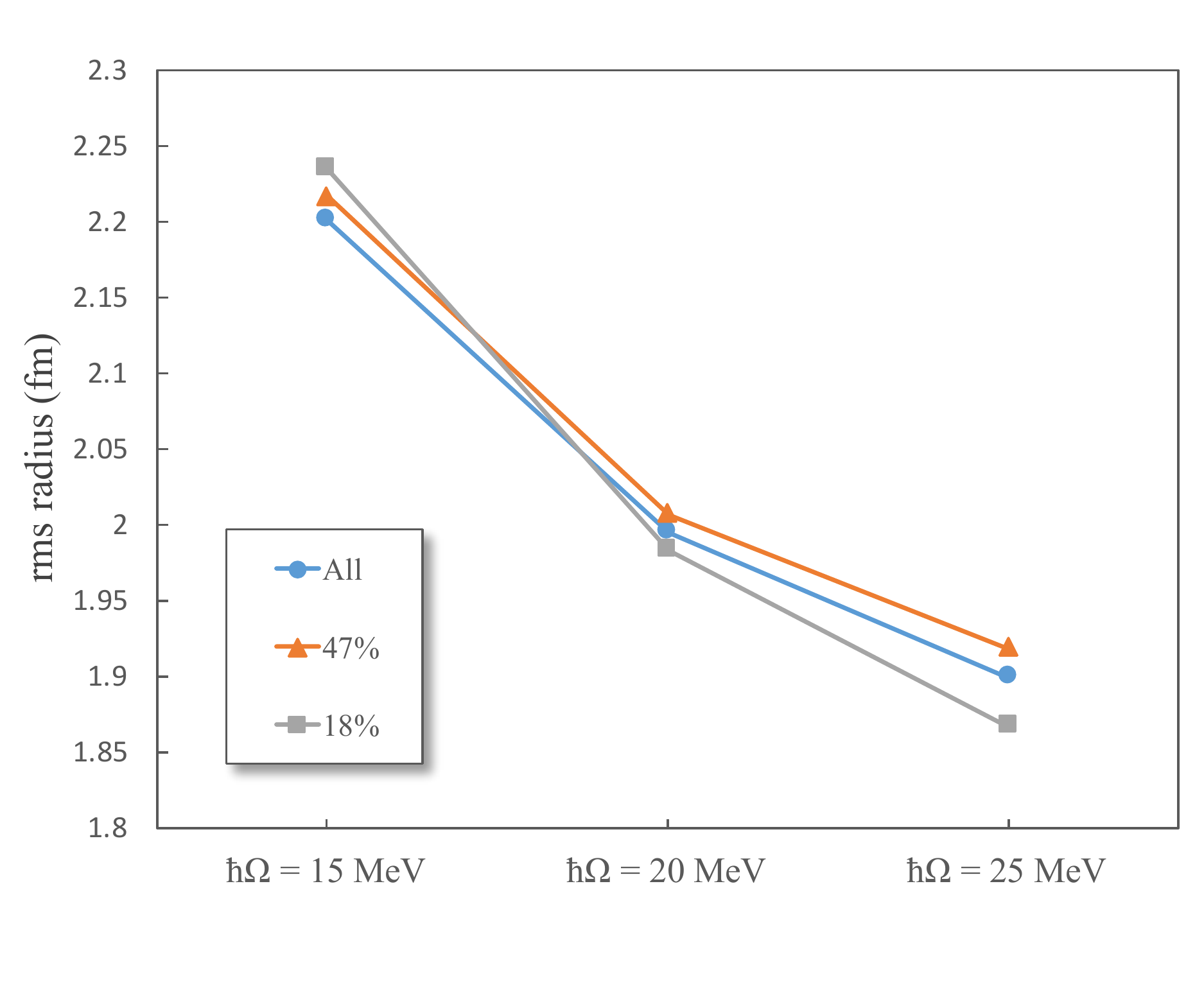}
\caption{\label{fig:rms_select_hw} $^{12}$C  \gs rms radius from SA-NCSM calculations with \Nmax= 6 model space vs. $\hbar\Omega$, using the  full (``All") and selected (labeled by the percentage of the tensors kept) JISP16 interaction.  
}
\end{figure}

\begin{figure} 
\includegraphics[width=0.5 \textwidth]{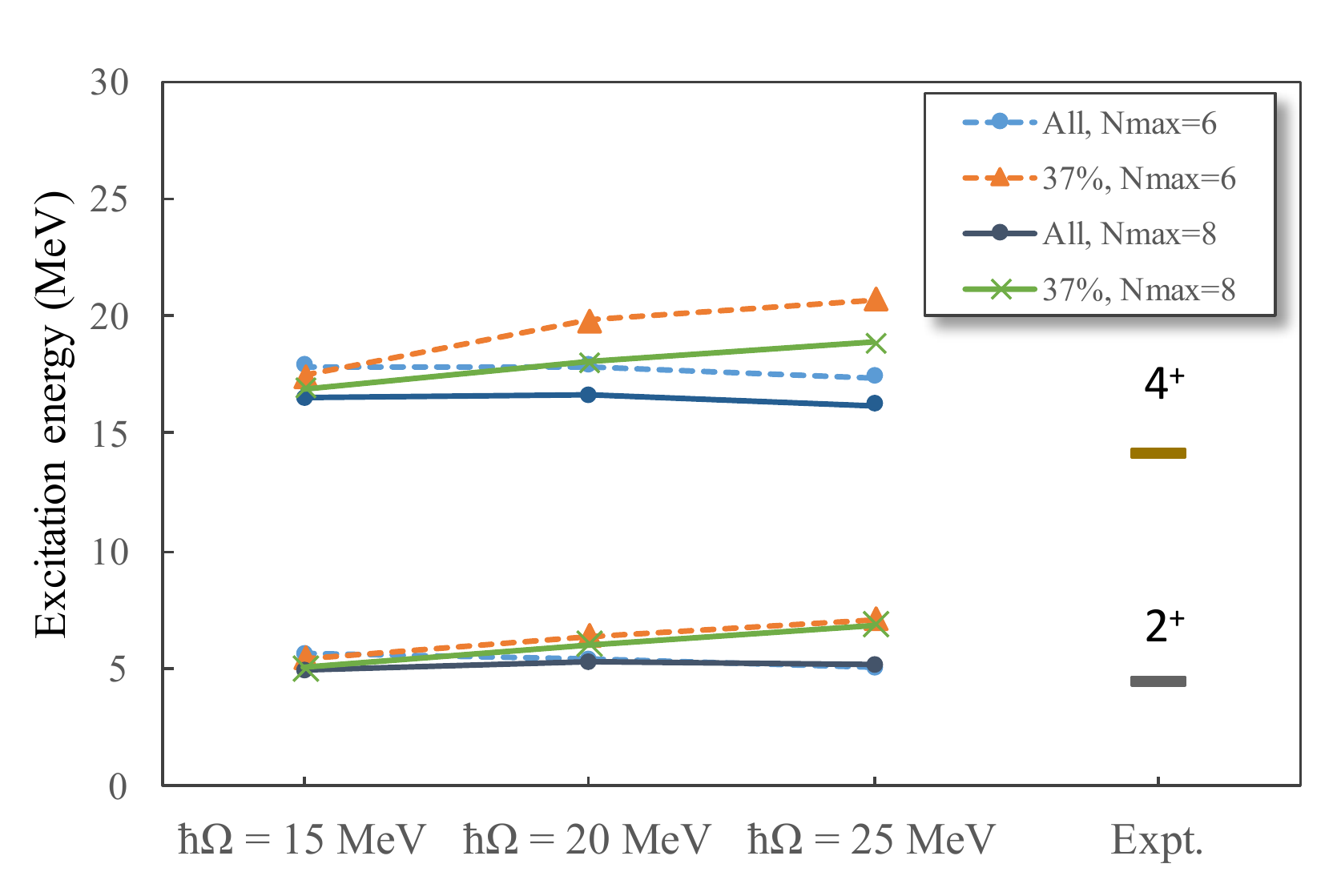}
\caption{\label{fig:en_hw}  Excitation energies  of the first $2^+$ and $4^+$ states for $^{12}$C from SA-NCSM calculations with \Nmax= 6 and \Nmax= 8 model spaces using full  JISP16 interaction (``All") and its selected counterpart (with 37\% of the tensors kept), with $\hbar\Omega=15, 20 $ and $25$ MeV, and compared to experiment.
}
\end{figure}

As mentioned above, the dependence on the HO parameter $\hbar\Omega$   disappears at the \Nmax$\rightarrow \infty$ limit, however, even for comparatively small \Nmax model spaces, there is often a range of $\hbar\Omega$ values, which achieves convergence for selected observables, while typically larger \Nmax model spaces are required  outside  this range. For long-range observables, such a range often  falls closely to an empirical estimate given by $\hbar\Omega=41/A^{1/3}$ \cite{BohrMottelson69}, which is 18 MeV for $^{12}$C. We investigate the dependence of the \gs rms radius of $^{12}$C on  $\hbar\Omega$ using different selections  (Fig. \ref{fig:rms_select_hw}). We examine  small model spaces, where the $\hbar\Omega$ dependence is large and its effect on the interaction selections is expected to be enhanced; yet, we ensure that these model spaces provide results close to the \Nmax=12 outcomes  (see \Nmax=6 and 8 results in Figs. \ref{fig:12Cen} and \ref{fig:obs}).  Comparing to the full interaction, the results indicate that, indeed, small deviations are observed for values around $\hbar\Omega=18$ MeV, and the deviations become larger  at higher (less optimal) $\hbar\Omega$ values (Fig. \ref{fig:rms_select_hw}).
Similarly,  the excitation energies for $\hbar\Omega = 15$ MeV calculations are much less sensitive to the interaction selection (Fig. \ref{fig:en_hw}),whereas the deviation in the results between the initial and selected interactions increases for higher  $\hbar\Omega$. However, this difference gets smaller with increasing model space. To summarize, the selection of the interactions affects the calculations with optimal $\hbar\Omega$ values the least. 

\begin{figure} 
\includegraphics[width=0.49\textwidth]{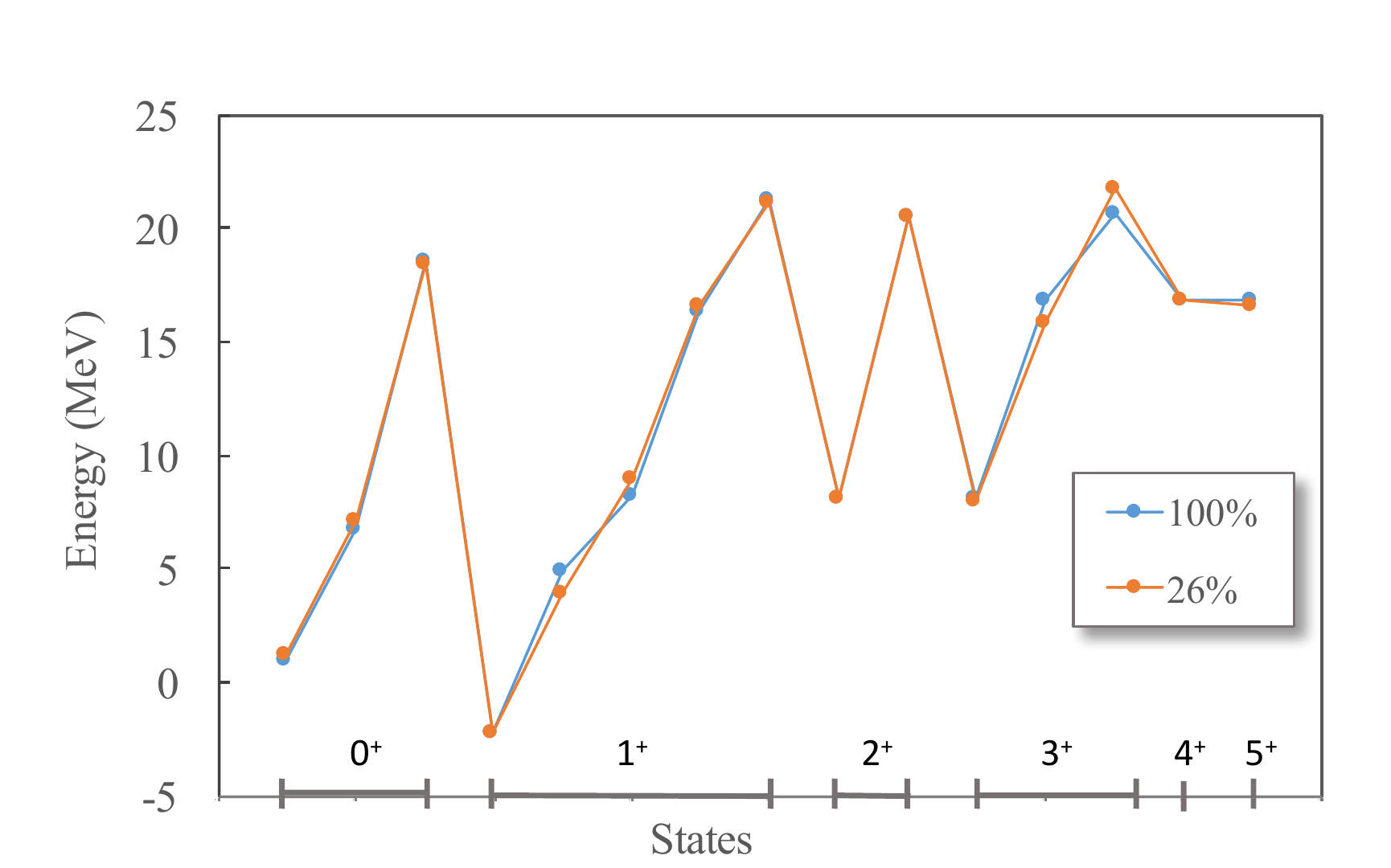}
\caption{\label{fig:2H} Energies of the proton-neutron system for the  positive-parity lowest-lying states ($<$ 30 MeV), calculated in the SA-NCSM in \Nmax=12 model space using the JISP16 interaction, with all  terms kept  (100\%) as compared to a selection that keeps only 26\% of the terms, for $\hbar\Omega$=15 MeV.
}
\end{figure}

It is interesting to examine how the selection of NN interactions affects the nucleon-nucleon physics. As a simple illustration, we  study the Hamiltonian for the proton-neutron system and its corresponding eigenvalues. 
In addition to $T=0$ states, we consider $T=1$ states, which can also inform the proton-proton and neutron-neutron systems. To do this, we look for deviations in  the corresponding eigenvalues as compared to those computed with the full interaction. We note that these comparisons use bound single-particle basis states,  so results will not apply to the proton-neutron scattering states,  however, using the same many-body method, any deviations will inform about the interaction selection. In particular, we observe that only about quarter of the \SU{3}-symmetric interaction components (the most dominant ones) can reproduce, with high accuracy, the energies that use the full interaction for most of the low-lying states of the proton-neutron system  (Fig. \ref{fig:2H}). To estimate the difference in energies, we calculate the root mean square error (RMSE),
$\mathrm{RMSE} = \sqrt{\frac{1}{N_d}\sum_i^{N_d} \Big(E_{\rm all}^i-E_{\rm sel}^i\Big)^2}$
 where $E_{\rm all}$ and $E_{\rm sel}$ are the eigenenergies calculated with the initial and selected interactions, respectively, the summation is over all positive- or negative-parity states and $N_d$ is the total number of states.
For negative-parity $0\le J \le 5$ states up through energy with 30 MeV, we find RMSE to be about 0.9 - 1.2 MeV  depending on $\hbar\Omega$, whereas for positive-parity states, it is between 0.5 and 0.9 MeV. Similar RMSE values are seen even for the higher lying spectrum up to 50 MeV.
As it can be seen from Fig. \ref{fig:2H}, the main deviations come from the second and third $1^+$ and $3^+$ states indicating that certain states are more sensitive to the selection than others. 

\subsection{Dominant features in realistic interactions}

There are various techniques of renormalization such as OLS and SRG that are employed to ``soften'' the realistic interactions, which in turn can be used in comparatively smaller model space. Comparing the \SU{3} decompositions of initial interactions to their renormalized (effective) counterparts shows that the same major \SU{3} tensors remain dominant after renormalization (Fig. \ref{fig:decomp}). In the case of JISP16 the tensors with the largest relative strengths practically do not change. 
The renormalization has a larger impact on the N3LO interaction where  the spread over various tensors is larger.
Here, only a few SU(3)-symmetric components change significantly while the others change slightly. It should be noted that the two effective counterparts of the interactions resemble each other (Fig. \ref{fig:decomp}). A similar behavior is observed for, e.g., the
AV18 \cite{WiringaSS95} and CD-Bonn interactions \cite{LauneyDD16}.

Examining the largest contributing tensors of realistic interactions we can link them to the monopole operator (the HO potential), $Q \cdot Q$, pairing, spin-orbit and tensor forces. The key idea is that the position and momentum operators, $\vec r$ and $\vec p$ respectively, have an \SU{3} rank $(1\,0)$, and conjugate $(0\,1)$ (to preserve hermicity), with SU(2)$_S$ rank zero ($S_0=0$, that is, the operator does not change spin).  Hence,  the HO potential operator ($\sim r^2=\vec r \cdot \vec r$) has  orbital momentum $L_0=0$ and spin $S_0=0$, and \SU{3} rank of $(2\,0)$ and  $(0\,0)$ (and conjugates), whereas the quadrupole operator $Q$, given by the tensor product of $\vec r$, has $L_0=2$ and  $S_0=0$, and  $(2\,0)$ and  $(1\,1)$ (and conjugates)  \cite{TobinFLDDB14}; similarly for the tensor force, but with $L_0=2$ and  $S_0=2$ . The $Q \cdot Q$ operator, which describes the interaction of each nucleon with the quadrupole moment of the nucleus, will then have $L_0=0$ and spin $S_0=0$, along with $(4\,0)$, $(2\,0)$, $(2\,2)$ and  $(0\,0)$  (and conjugates). The spin-orbit operator has $L_0=1$ and  $S_0=1$, with an \SU{3} rank of $(1\,1)$. Indeed, the scalar (0 0) $S_0$=0 dominates for a variety of realistic interactions, and especially in their effective counterparts (see Fig. \ref{fig:decomp}); it is typically followed by 
 $(2\, 0)$, $(4\, 0)$ and (2 2)$S_0$=0 and their conjugates. These \SU{3} modes are the ones that appear in the $Q \cdot Q$ interaction, while $(\lambda \, \lambda)$ configurations dominate the   pairing interactions within a shell \cite{BahriED95}. 
The dominant $(2\,0)$ and $(1\,1) S_0=2$ modes, and conjugates,
can be linked to the tensor force. Finally, the $(1\, 1)S_0$=1 can be linked to the spin-orbit force. These features, we find, repeat for various realistic  interactions and, more notably, the similarity is found to be further enhanced for their renormalized counterparts.
Given the link between the phenomenon-tailored interactions and major terms in realistic interactions, it is then not surprising that both \emph{ab initio} approaches and earlier schematic models can successfully describe dominant features in nuclei.

\section{Conclusions}

Realistic NN interactions expressed in \SU{3} basis show a clear dominance of only a small fraction of terms. 
We performed \emph{ab initio} calculations of several observables in $^{12}$C using interactions that were selected down to the most significant terms and compared them to the calculations with the initial interactions. We found that for the small $\hbar\Omega$ values even the interactions with less than half of the terms produce almost the same results as the initial interaction for the low-lying spectrum, B(E2) values and rms radii of $^{12}$C. The selection appears to affect more the calculations that use interactions with higher $\hbar\Omega$ values in small model spaces, however the deviations between the initial and selected interaction results decrease as the model space becomes larger. In addition, the eigenvalues of the proton-neutron system for all of the positive and negative parity states below 30 MeV change only slightly with as few as the quarter of the initial interaction terms.  

By analyzing the most dominant terms of various realistic interactions, we found that they can be linked to well known nuclear forces. In particular, inspection of these terms allowed us to link them to the widely used HO potential, $Q \cdot Q$, pairing, spin-orbit and tensor forces. Moreover, we saw that after renormalization the NN interactions, regardless of their type, have mainly the same dominant terms with similar strengths, indicating that the renormalization techniques strengthen the same dominant terms in all interactions. 

\section*{ACKNOWLEDGMENTS}
Support from the U.S. National Science Foundation (ACI -1713690, OIA-1738287, PHY-1913728), the Czech Science Foundation (16-16772S) and the Southeastern Universities Research Association are all gratefully acknowledged. This work benefitted from computing resources provided by NSF's Blue Waters computing system (NCSA), LSU  (www.hpc.lsu.edu), and the National Energy Research Scientific Computing Center (NERSC). 

\section*{APPENDIX}
\label{apdx}
In standard second quantized form, a one- and two-body interaction
Hamiltonian is given
in terms of fermion creation $a_{ jm(1/2)\sigma }^\dagger$ and annihilation
$\tilde{a}_{ j-m(1/2)-\sigma } = (-1)^{j-m+1/2 -\sigma }a_{ jm(1/2)\sigma }$
tensors, which create or annihilate a particle of type
$\sigma =\pm 1/2$ (proton/neutron) in the HO basis. 

In Eq. (\ref{V2ndQF}),
$V_{rstu}^{\Gamma}$ is the two-body antisymmetric 
matrix element in the
$JT$-coupled scheme [$V_{rstu}^{\Gamma
}=-(-)^{r+s-\Gamma}V_{srtu}^{\Gamma }=
-(-)^{t+u-\Gamma } V_{rsut}^{\Gamma }=(-)^{r+s-t-u}V_{srut}^{\Gamma }=
V_{turs}^{\Gamma }$]. For an isospin nonconserving two-body
interaction of isospin
rank ${\mathcal T} $, the coupling of fermion operators is as follows,
$\{\{a_r^\dagger \otimes
a_s^\dagger\}^{JT}\otimes \{a_t \otimes a_u\}^{JT} \}^{(0{\mathcal T})}$, with
$V_{rstu}^{({\mathcal T}) J T}$ matrix elements.
\begin{widetext}
\begin{eqnarray}
V&=&-\frac{1}{4}\sum_{rstu \Gamma}
\sqrt{(1+\delta _{rs})(1+\delta _{tu})}\Pi_\Gamma V_{rstu}^\Gamma
\{\{a_r^\dagger \otimes a_s^\dagger\}^\Gamma \otimes
\{\tilde a_t \otimes \tilde a_u\}^\Gamma \}^{(\Gamma_0 M_{\Gamma_0})} \nonumber\\
&=&
\sum_{{\tiny 
\begin{array}{c}
   (\chi^* \omega S)_{fi}    \\
   \rho_0 \omega_0 \kappa_0 S_0
\end{array}
}}
\frac{(-1)^{\omega_0-\omega_f+\omega_i}}{\sqrt{(1+\delta_{\eta_r\eta_s})(1+\delta_{\eta_t\eta_u})}}
\frac{1}{\Pi_{S_0}}\sqrt{\frac{\dim \omega_f}{\dim \omega_0}}
V_{(\chi \omega S)_{f,i}T}^{\rho_0 \omega_0 \kappa_0 S_0} \times \nonumber\\
&&\sum_{\rho'_0} \Phi_{\rho'_0 \rho_0}(\omega_0 \omega_i \omega_f)
\{\{a_{\eta_r}^\dagger \otimes a_{\eta_s}^\dagger\}^{\omega_f S_f T} \otimes
\{\tilde a_{\eta_t} \otimes \tilde a_{\eta_u}\}^{\omega_i S_i T} \}^{\rho'_0 \omega_0 \kappa_0 (L_0=S_0 S_0)\Gamma_0=0 M_{\Gamma_0}=0} ,
\label{V2ndQF}
\end{eqnarray}

\end{widetext}

where dim $\omega$ is defined in Eq. \ref{dim} and the phase matrix $\Phi_{\rho'_0 \rho_0}(\omega_0\omega_i\omega_i)$ accommodates the interchange between the coupling of $\omega_0$ and $\omega_i$ to $\omega_f$, so for \SU{3} Clebsch-Gordan coefficients we have \cite{Escher1997thesis}

\begin{widetext}
\begin{equation}
  \CG{\omega_0\kappa_0 L_0 M_0}{\omega_i \kappa_i L_i M_i }{\omega_f\kappa_f L_f M_f}_{\rho_0} = \sum_{\rho'_0} \Phi_{\rho_0\rho'_0}(\omega_0\omega_i\omega_f)\CG{\omega_i\kappa_i L_i M_i}{\omega_0 \kappa_0 L_0 M_0 }{\omega_f\kappa_f L_f M_f}_{\rho'_0}.
\end{equation}

\end{widetext}

For the special case when $\rho=1$, that is, where the \SU{3} coupling $\{\omega_i \otimes\omega_0\} \rightarrow\omega_f$ is unique, the phase matrix reduces to a simple phase factor $(-1)^{(\lambda_0+\mu_0)+(\lambda_i+\mu_i)-(\lambda_f+\mu_f)}$.  Finally, the interaction reduced matrix elements in  a $\SU{3}\times\SU{2}_S\times\SU{2}_T$-coupled HO basis are given as,
\begin{widetext}
\begin{eqnarray}
V_{(\chi \omega S)_{fi};T}^{\rho_0 \omega_0 \kappa_0 S_0} &=& (-)^{S_f+S_0} \Pi_{TS_0} \frac{ \dim \omega_0}{\dim \omega_f} 
\sum_{J(\kappa L)_{if}} (-)^{L_i+J} \Pi_J^2 \Pi_{L_f} \Wigsixj{L_f}{S_f}{J}{S_i}{L_i}{S_0} \RedCGw{i}{0}{f}{0} V^{\Gamma}_{(\chi \omega \kappa L S)_{fi}}
\nonumber \\
& =&
(-)^{S_f+S_0}\Pi_{TS_0} \frac{\dim \omega_0}{\dim \omega_f} 
\sum_{{\tiny 
J(\kappa L)_{if}
}}
(-)^{L_i+J} \Pi_J^2 \Pi_{L_f} \Wigsixj{L_f}{S_f}{J}{S_i}{L_i}{S_0} \RedCGw{i}{0}{f}{0} \times \nonumber \\
&&
\Pi_{L_iL_fS_iS_f}
\sum_{{\tiny 
\begin{array}{c}
    l_rl_sl_tl_u    \\
    j_rj_sj_tj_u 
\end{array}
}}
 \sqrt{\frac{(1+\delta_{rs})(1+\delta_{tu})}
 {(1+\delta_{\eta_{r}\eta_{s}})(1+\delta_{\eta_{t}\eta_{u}})}} \Pi_{j_{r}j_{s}j_{t}j_{u}}
 \RedCG{(\eta_{r}\,0) l_{r}}{(\eta_{s}\,0)l_{s}}{(\omega \kappa L)_f} \times \nonumber \\
&&
 \RedCG{(\eta_{t}\,0)l_{t}}{(\eta_{u}\,0)l_{u}}{(\omega \kappa L)_i} 
 \Wigninej{ l_{r} }{\half}{ j_{r} }{ l_{s} }{\half}{ j_{s} } {L_f}{S_f}{J}
 \Wigninej{ l_{t} }{\half}{ j_{t} }{ l_{u} }{\half}{ j_{u} } {L_i}{S_i}{J}
 V^\Gamma_{rstu},
 \label{JTtoSU3}
\end{eqnarray}
\end{widetext}
where $V^{\Gamma}_{(\chi \omega \kappa L S)_{fi}}$ is a two-body interaction in a \SU{3}-$JT$-coupled scheme, as mentioned above $\RedCG{}{}{}$ are reduced SU(3) Clebsch-Gordan coefficients, and we use \SU{2} Wigner 6-j and 9-j symbols.

\bibliographystyle{apsrev}
\bibliography{SU3-guided_int.bib}

\end{document}